\documentclass[aps,prl,preprint]{revtex4}
\usepackage{graphicx}%
\usepackage{citesort}

\def\HfO2{{HfO$_2$}}
\def\SiO2{{SiO$_2$}}

\def\cm1{{cm$^{-1}$}}

\def\fig#1{{Fig.~\ref{fig:#1}}}

\begin{document}

\title{Building effective models from sparse but precise data}

\author{Eric Cockayne}
\affiliation{Ceramics Division, Materials Science and
Engineering Laboratory, National Institute of Standards and
Technology, Gaithersburg, Maryland 20899-8520}

\author{Axel van de Walle}

\affiliation{Engineering \& Applied Science Division, California Institute of Technology, Pasadena, CA 91125}

\date{\today}

\begin{abstract}

A common approach in computational science is to use a set of of highly
precise but expensive calculations to parameterize a model that allows less
precise, but more rapid calculations on larger scale systems.  
Least-squares fitting on a model that underfits the data is generally used for this purpose.  
For arbitrarily precise data free from statistic noise, {\it e.g.} {\it ab initio} calculations, 
we argue that it is more appropriate to begin with a ensemble of models that overfit the data.  
Within a Bayesian framework, a most likely model can be defined that incorporates physical
knowledge, provides error estimates for systems not included in the fit, and reproduces
the original data {\em exactly}.  We apply this approach to obtain a cluster expansion model for the
CaZr$_{1-x}$Ti$_x$O$_3$ solid solution.

\end{abstract}

\maketitle

A common approach in computational science is the
use of a small number of highly precise but expensive calculations to generate data
to fit the parameters of a less accurate but more computationally tractable ``effective model''
enabling larger-scale simulations.\cite{Yip05,Olson97}
A typical example is the fit of a simplified energy model
to accurate quantum mechanical calculations.\cite{Hart05,fontaine:clusapp,ducastelle:book,
Connolly83,asta:fppheq,ceder:oxides,zunger:NATO,vandeWalle02,Ruban08}
Although least-squares minimization is traditionally used for this purpose,
it is not commonly recognized that this approach implicitly and incorrectly assumes
that the uncertainty lies in the data rather than in the effective model.

Here we show that the fact that the model is less accurate than the data
can be properly taken into account within a Bayesian\cite{Jaynes:prob} framework where
the ``prior'' probability distribution of the model parameters characterizes the range of physically plausible models.
The model parameters are obtained by maximizing the ``posterior'' distribution provided by Bayes rule, given the accurate data and the prior.
This approach enables a perfect fit to the input noiseless data, while avoiding the
usual artifacts of overfitting\cite{vandeWalle02}
and enables the seamless inclusion of physical knowledge into the fitting procedure via the prior.
Although Bayesian methods have a long history in the statistical sciences, 
(including recent interest in Bayesian learning 
techniques\cite{Tipping01,Figueiredo02}),
the unexpectedly well-behaved limit of
completely noiseless data we report here has, to our knowledge, not been noted,
perhaps because existing methods have historically been motivated by the need to fit noisy experimental data
rather than noiseless calculated data. However, the latter setting clearly deserves more attention.

While our general theoretical approach should have broad applicability in numerous
fields of computational sciences, in this Letter, we focus on
the specific but broadly applicable example of the construction of an efficient
energy model for a crystalline alloy.\cite{Hart05}
This task has immediate applications to thermodynamic modeling of alloys and the determination of
their phase diagrams, a crucial component of alloy design and optimization.

In this context, the accurate total energy data are provided by
{\it ab initio} electronic structure methods based upon density functional theory (DFT),\cite{jones:ldareview,Payne:Pseudo-review}
whose accuracy has been thoroughly validated in a wide range of solid-state systems.\cite{perdew:appgga}
(Although such {\it ab initio} calculations may not provide the exact quantum mechanical result,
they are precise in that they are virtually free of random errors, as numerical noise
is well-controlled in modern {\it ab initio} software.)
The effective model is a so-called {\it cluster expansion} (CE),\cite{Sanchez84,ducastelle:book,fontaine:clusapp,Connolly83,Hart05,asta:fppheq,ceder:oxides,zunger:NATO,vandeWalle02,Ruban08} that takes the form
of a polynomial in occupation variables (described in detail below) indicating which atom lies on each lattice site.
The unknown parameters of CE to be determined are the coefficients of this polynomial.
The CE has been previously shown \cite{Sanchez84} to be able, in principle, to exactly represent any possible
configurational-dependence of the energy, provided that {\it all} terms the expansion 
are included, which unfortunately amounts to an infinite number of terms.

Typically, such CE models are created through a least squares fit
to a database of {\it ab initio} structural energies obtained using a CE
truncated to a finite number of terms, so that the number of input configurations
is larger than the number of unknown parameters. This 
leads to a ``truncation problem", where the
terms to be retained in the model must be determined.
Approaches for optimizing the truncation in this context
have included the cross-validation score minimization\cite{vandeWalle02,Hart05,Sluiter96},
sometimes combined with regularization techniques\cite{Drautz07,Zunger04}
and conventional Bayesian approaches.\cite{Jansen08,Mueller09}

These approaches treat systematic errors (due to model truncation)
and statistical errors (due to numerical noise in the data) on an equal footing
without exploiting the knowledge that statistical errors are, in fact, negligible in this context.
In the large sample limit, truncation selection methods would eventually ``discover'' that
the statistical noise is zero, but considerable improvements are possible if this known fact
is explicitly taken into account from the start.

In this Letter, we avoid truncation problems by including many more
terms in the effective model than the number of {\it ab initio}
calculations performed. Although the fitting problem is
underdetermined, it can nonetheless be solved by using
Bayesian inference with a physically based prior probability distribution 
for the model coefficients.

A configuration $i$ in a binary alloy is defined
by the occupation of each site $k$ of a lattice by one of two
species, indicated by a spin-like variable $s_{ik}=\pm 1$.
Each configuration $i$ has an associated {\it ab initio}
energy $E_i$.  These energies can be written in terms of
effective cluster interactions\cite{Sanchez84,Connolly83}: 
\[
\sum_\alpha \xi_{i\alpha} J_{\alpha} = E_i,
\]
where $\xi_{i\alpha}=\left\langle \prod_{k \in \alpha} s_{ik} \right\rangle$ is the translationally and rotationally
averaged multibody spin correlation for each symmetry-independent
cluster $\alpha$, while $J_{\alpha}$ is the associated effective interaction parameter to be determined.

\begin{figure}[h]
\includegraphics[width=144mm]{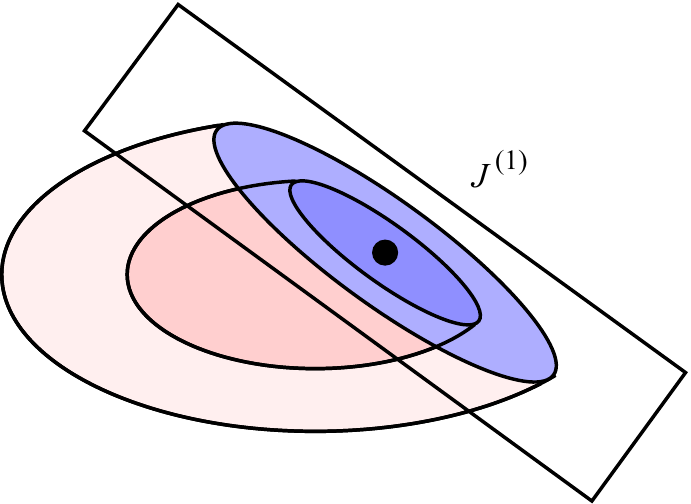}
\caption{(Color online) The ellipsoids schematically represent equiprobability surfaces of
a many-dimensional prior probability distribution for $P^{0}(J)$.
The plane represents the constraints on J given by the results of an {\it ab initio} total
energy calculation. The $(N-1)$-dimensional ellipsoid sliced by the plane gives an
equiprobability surface for the posterior distribution of $P^{(1)}(J)$;
the point marked $J^{(1)}$, represents the most likely solution for $J$.}
\label{fig:post}
\end{figure}

Bayes' theorem\cite{Jaynes:prob} states that given a prior probability
distribution $P^{(0)}(J)$ of the unknown parameter vector $J$ and one energy observation $E_i$, the
posterior probability of $J$ given $E_i$ is
\[
P(J|E_i) \propto P(E_i|J) P^{(0)}(J).
\]
Since $E_i$ is precisely known, the conditional probability reduces to a delta function
\begin{equation}
P(E_i|J) = \delta(\xi_i J - E_i),
\label{eqndelta}
\end{equation}
where $\xi_i$ denotes the row vector of all values of $\xi_{i\alpha}$ for a fixed $i$.
As a result, $P(J|E_i)$ is trivially proportional to
$\delta(\xi_i J - E_i) P^{(0)}(J)$.
By induction, the posterior probability of $J$ based on all the
energy information $E_i$, $i = 1,\ldots,n$ is
\[
P^{(n)} (J) \propto  \prod_{i=1}^{n} \delta(\xi_i J - E_i) P^{(0)}(J).
\]
The difference between prior and posterior
probabilities is represented geometrically in \fig{post}.
Each new data point selects a cross-section of the prior distribution corresponding to sets of
parameter values that agree perfectly with this data point. Within the intersection of these cross-sections, each point has a different posterior
probability that is dictated by the prior.
The most likely model parameters $J^{(n)}$ can be determined by 
maximizing $P^{(n)} (J)$ (see Key methodological details, at end).
This approach selects a unique solution from an otherwise underdetermined system of equations,
based on the ``physical'' information provided by the prior.

The width of the posterior provides a measure of the uncertainty remaining in the fitted parameter after the data has been incorporated,
which can be used, for instance, to access the accuracy of predicted energies for any structure not included in the fit.
For a Gaussian prior, the posterior is Gaussian as well and the most likely parameter values are also the expected parameter values, which
implies, given the linearity of the cluster expansion, that the predicted energies from the CE model will also be expected values.

\begin{figure}[h]
\includegraphics[width=144mm]{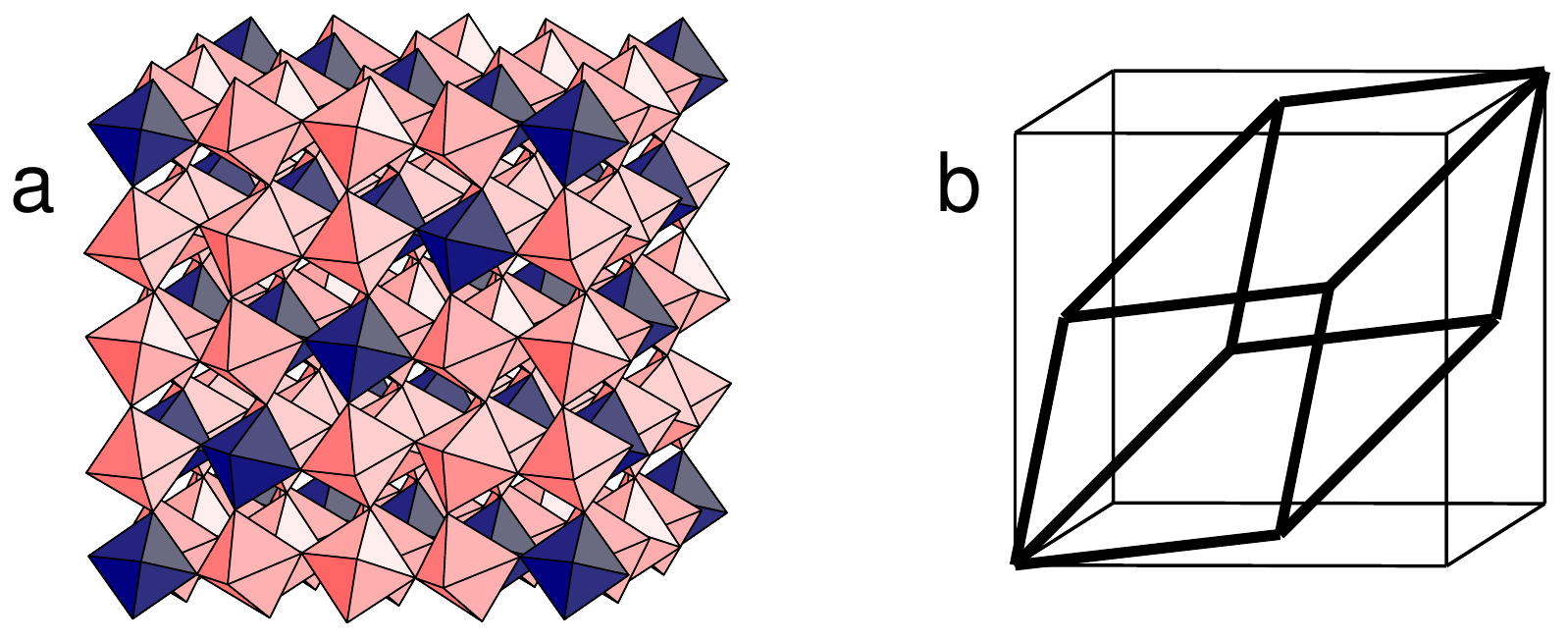}
\caption{(Color online) (a) Representative CZT structure; (b) Common 80-atom supercell for CZT energy calculations
highlighted in bold.}
\label{fig:cell}
\end{figure}

The above procedure was applied to modeling total energies in 
the CaZr$_{1-x}$Ti$_x$O$_3$ (CZT) system, a perovskite solid solution with tilted oxygen 
octahedra.\cite{Levin06}
The structures studied were constrained to a common
80-atom supercell shown in \fig{cell}, which has 16 perovskite ``B'' sites that contain
either Zr or Ti.
Local density functional theory calculations were
performed on specific CZT configurations using the
\textsc{VASP}\cite{Kresse99}
code and ultrasoft pseudopotentials,\cite{Vanderbilt:soft_pseudo},
with semicore p electrons treated as valence electrons for Ca, Zr, and Ti.
A 375 eV plane wave energy cutoff and a 1500 eV cutoff for augmentation charges were used.
The $k$-point mesh was equivalent to an $8\times 8\times 8$ Monkhorst-Pack
grid for a primitive perovskite cell.
The number of symmetrically distinct possible arrangements of Zr and Ti and the number of 
terms in the full cluster expansion are both 2386; {\em all} terms are retained in
the fitting.

Based on simple physical considerations, the following Gaussian prior was selected
\[
P^{(0)} (J) = \prod_{\alpha} (\sqrt{2 \pi} w_\alpha)^{-1} \exp \left((-J_\alpha)^2/(2 w_\alpha^2)\right),
\]
with
$w_\alpha  = 
A b^{n_\alpha} \prod_{\{i,j\} \subset \alpha} (r_{ij}/r_{\rm min})^{-2}$.
Clusters with more sites $n_\alpha$ are expected to have smaller
coefficients ({\it i.e.} $b < 1$), and clusters with pairs $\{i,j\}$ of atoms at larger
separations $r_{ij}$ are expected to have smaller coefficients.
The exponent of $-2$ is motivated by the observation that Zr and Ti
have the same charge, so interactions cannot be mediated by differences
in monopole coupling, leaving only dipolar leading terms.
We set $b = 0.17$, based on the ratio of the quadratic to linear terms
in a Taylor expansion to the energies of three trial structures 
(pure CaZrO$_3$, pure CaTiO$_3$, and rocksalt-ordered 
CaZr$_{1/2}$Ti$_{1/2}$O$_3$), and $A =$ 1.58 eV per atom, 
based on scaling the error estimates for all 30 structures included 
in the final fit (see Key methodological details).

\begin{figure}[h]
\includegraphics[width=144mm]{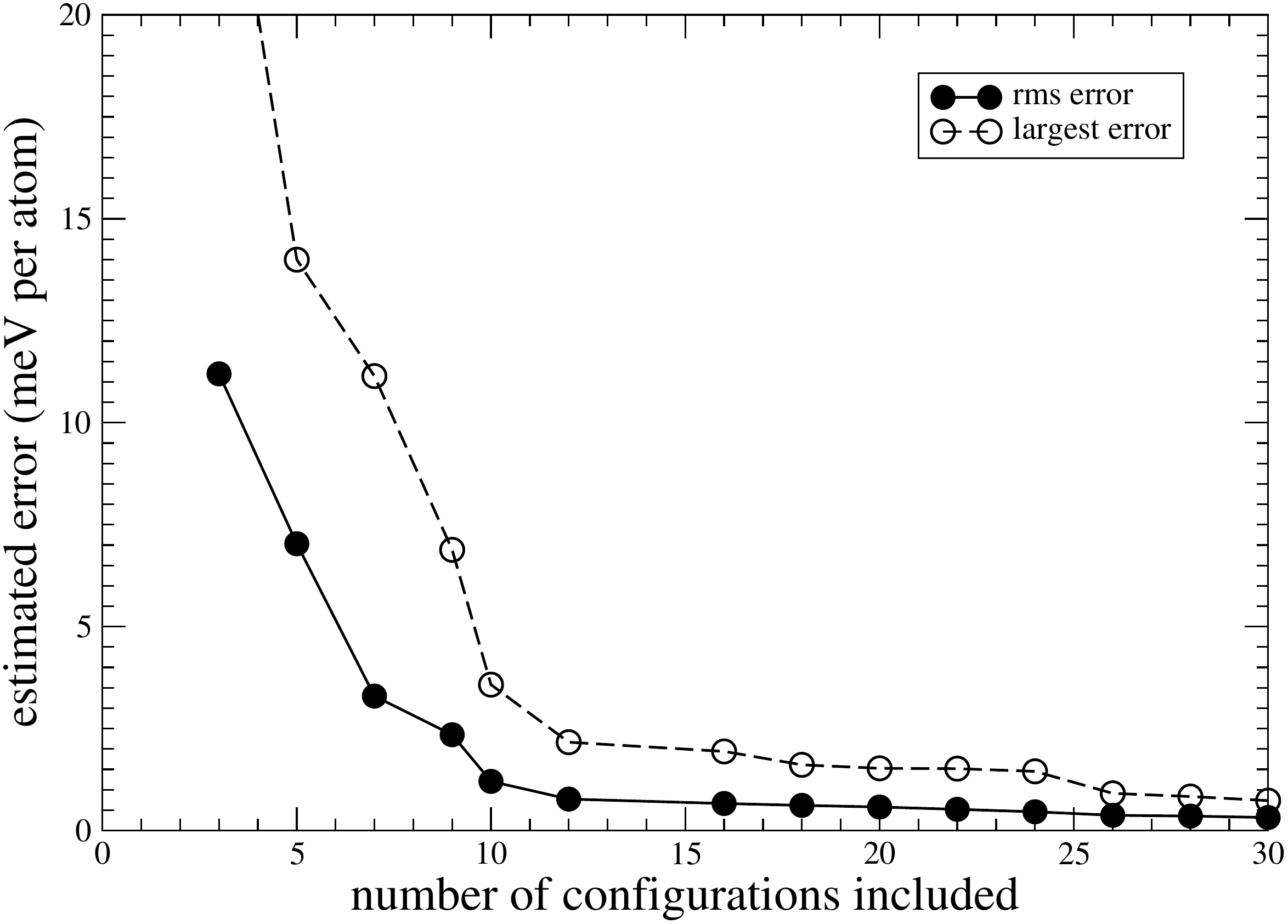}
\caption{Convergence of the mean and maximum predicted errors on energies of
configurations not included in the fit, as a function of
the number of configurations included in fit.}
\label{fig:errors}
\end{figure}

Our Bayesian approach is embedded in an iterative scheme where,
at each step, a most likely solution $J$ is found and error estimates for
the other configurations not included in the fit are calculated.
The configuration with the highest estimated error
is then added to the fit, after its {\it ab initio} total
energy is calculated (see Key methodological details). 
This procedure is repeated until a sufficient predictive accuracy
has been reached, as illustrated in \fig{errors}.
A monotonic, rapid improvement of the fit is found as the
number of structures included in the fit increases.
For calculation efficiency purposes, one may truncate
the terms to be retained in the solution to those
whose $J_\alpha$ exceeds some small threshold value,
with little effect on the 
results (as suggested in \cite{Drautz07} in a related context); 
in our methodology, these terms can be retained in estimating errors 
on predicted energies.

For thermodynamic applications, such as phase diagram calculations,\cite{vandeWalle02}
it is crucial that the lowest lying energies have the proper order and energy differences.
The least-squares fitting procedure does not guarantee this,
{\em even if these states were included in the fitting procedure}.
By reproducing the energies of all states included
in the fit exactly, our procedure avoids this problem.

While the need to specify a prior in Bayesian methods is often criticized,
it should be realized that a conventional least-squares fit is not free of {\it a priori} assumptions either.
In the cluster expansion example, a conventional fit with a user-specified truncation distance amounts to a prior which is uninformative (flat) for the
included interactions coefficients but entirely concentrated at zero for the excluded coefficients.
This incorporates physical knowledge into the problem,
but with a complete certainty that far exceeds what a researcher could plausibly know.
A smoother prior which gradually concentrates the probability towards zero coefficient values as
the range of interaction increases appears a more appropriate description of a priori information.

If the data does contain some error (for instance, significant numerical noise), then the delta function in 
(\ref{eqndelta})
must be replaced by a smooth density and a conventional Bayesian methodology would result (as in \cite{Jansen08,Mueller09}). If that density is Gaussian,
then one recovers a so-called ridge regression or Tikhonov regularization\cite{tikhonov:reg}, a penalized least-squares estimator,
which has been previously used in the context of CE construction.\cite{Zunger04,Drautz07}

One possible way to select a prior is to estimate a range of likely
coefficient values on the basis of analogous systems.
Good theoretical knowledge of the physics governing the interactions is also useful.\cite{Mueller09}
Another possibility is to use the data itself to optimize the
choice of the prior.  Although this is not appropriate in a strict Bayesian context,
it is conceivable that some type of
cross-validation approach could be a valid prior-selection method
in a hybrid Bayesian/frequentist context.
This scheme is the noiseless limit of an existing variational cross-validation method\cite{Drautz07}.
Our preliminary results in this vein indicate that the method does
have the ability to rule out clearly unphysical priors in favor
of more physically plausible priors.
Formally quantifying this method's discriminating power represents
a fruitful avenue of future investigation.

The construction of multibody interatomic force fields based upon quantum mechanical
data\cite{Sundu08,Gale03} represents an important possible application of the methodology proposed here. 
It differs from the alloy problem in that the energy is a function of a {\em continuous} set of
distance variables, rather than a discrete set of spacings on a lattice.
While force fields are often selected to be nonlinear in the unknown parameters,
energy models that are linear in the parameters (such as splines) are probably preferable in our context,
as they make the interaction optimization problem numerically stable and efficient.
The lack of physical basis for spline functions is alleviated by the possibility of including many more parameters
than data points and the ability to favor physically plausible interaction shapes via the prior.
Such an approach could lead to successful force fields models, with defined uncertainties 
and with broad applications in computational physics.

\textbf{Key methodological details:} With a zero-mean Gaussian prior for the vector $J$ containing all interactions $J_{\alpha}$, the posterior maximum is given by
$J^{(n)} = W \xi^{T}(\xi W \xi^{T})^{-1} E,$ where $W$ denotes the variance-covariance matrix of the prior while
$\xi$ denotes the matrix with elements $\xi_{i\alpha}$ and $E$ denotes the vector of all $n$ known structural energies $E_i$.
An equivalent approach is to perform the changes of variables
$\xi^{\prime} = \xi W^{1/2}$ and $J^{\prime(n)} = W^{-1/2} J^{(n)}$
(where we use the symmetric square root matrix), 
and find the minimum Euclidean norm solution $J^{\prime(n)}$ to $\xi^{\prime} J^{\prime(n)} = E$
using, for instance, standard singular value decomposition techniques.

The expected energy of any additional configuration $i=n+1$ (with correlations equal to the row vector $\xi_{(n+1)}$) is given by $\xi_{(n+1)} J^{(n)}$.
The root mean square error in this energy is simply given
by  $|{\rm Pr}^{\prime(n)}_{\perp}{\xi^{\prime}}_{n+1}|$,
where ${\rm Pr}^{\prime(n)}_{\perp}$ is a projector
in the prime space that simultaneously 
projects ${\xi^{\prime}}_{n+1}$ orthogonal
to all ${\xi^{\prime}}_{i \leq n}$.

If {\em all} terms $w_\alpha$ in a Gaussian prior are scaled by the same
value $x$, then: (1) $J^{(n)}$ remains unchanged; (2) all predicted errors
are scaled by $x$.   In an analogous way to the leave-one-out
cross validation method\cite{vandeWalle02},  one can take the set of structures
used to obtain $J^{(n)}$ , omit each single structure $i$ one-at-a-time,
and calculate the estimated and actual errors for
$E_i$ based on fitting the model to all the other structures.  
A self-consistent value for $x$ is then obtained by scaling the mean square estimated
errors to equal the actual errors.

\noindent\textbf{Acknowledgements} A. vdW. was supported by
the US National Science Foundation via TeraGrid resources at NCSA and
SDSC under grant TG-DMR050013N and by the US Department of Energy via grants
DE-FG07-071D14893 and DE-PS52-07NA28208.

\newpage


\begin{thebibliography}{29}
\expandafter\ifx\csname natexlab\endcsname\relax\def\natexlab#1{#1}\fi
\expandafter\ifx\csname bibnamefont\endcsname\relax
  \def\bibnamefont#1{#1}\fi
\expandafter\ifx\csname bibfnamefont\endcsname\relax
  \def\bibfnamefont#1{#1}\fi
\expandafter\ifx\csname citenamefont\endcsname\relax
  \def\citenamefont#1{#1}\fi
\expandafter\ifx\csname url\endcsname\relax
  \def\url#1{\texttt{#1}}\fi
\expandafter\ifx\csname urlprefix\endcsname\relax\def\urlprefix{URL }\fi
\providecommand{\bibinfo}[2]{#2}
\providecommand{\eprint}[2][]{\url{#2}}

\bibitem[{\citenamefont{Yip}(2005)}]{Yip05}
\bibinfo{author}{\bibfnamefont{S.}~\bibnamefont{Yip}},
  \emph{\bibinfo{title}{Handbook of Materials modeling}}
  (\bibinfo{publisher}{Springer}, \bibinfo{address}{The Netherlands},
  \bibinfo{year}{2005}).

\bibitem[{\citenamefont{Olson}(1997)}]{Olson97}
\bibinfo{author}{\bibfnamefont{G.}~\bibnamefont{Olson}},
  \bibinfo{journal}{Science} \textbf{\bibinfo{volume}{277}},
  \bibinfo{pages}{1237} (\bibinfo{year}{1997}).

\bibitem[{\citenamefont{Hart et~al.}(2005)\citenamefont{Hart, Blum, Walorski,
  and Zunger}}]{Hart05}
\bibinfo{author}{\bibfnamefont{G.~L.~W.} \bibnamefont{Hart}},
  \bibinfo{author}{\bibfnamefont{V.}~\bibnamefont{Blum}},
  \bibinfo{author}{\bibfnamefont{M.~J.} \bibnamefont{Walorski}},
  \bibnamefont{and} \bibinfo{author}{\bibfnamefont{A.}~\bibnamefont{Zunger}},
  \bibinfo{journal}{Nature Mater.} \textbf{\bibinfo{volume}{4}},
  \bibinfo{pages}{391} (\bibinfo{year}{2005}).

\bibitem[{\citenamefont{de~Fontaine}(1994)}]{fontaine:clusapp}
\bibinfo{author}{\bibfnamefont{D.}~\bibnamefont{de~Fontaine}},
  \bibinfo{journal}{Solid State Phys.} \textbf{\bibinfo{volume}{47}},
  \bibinfo{pages}{33} (\bibinfo{year}{1994}).

\bibitem[{\citenamefont{Ducastelle}(1991)}]{ducastelle:book}
\bibinfo{author}{\bibfnamefont{F.}~\bibnamefont{Ducastelle}},
  \emph{\bibinfo{title}{Order and Phase Stability in Alloys}}
  (\bibinfo{publisher}{Elsevier Science}, \bibinfo{address}{New York},
  \bibinfo{year}{1991}).

\bibitem[{\citenamefont{Connolly and Williams}(1983)}]{Connolly83}
\bibinfo{author}{\bibfnamefont{J.~W.} \bibnamefont{Connolly}} \bibnamefont{and}
  \bibinfo{author}{\bibfnamefont{A.~R.} \bibnamefont{Williams}},
  \bibinfo{journal}{Phys. Rev. B} \textbf{\bibinfo{volume}{27}},
  \bibinfo{pages}{5169} (\bibinfo{year}{1983}).

\bibitem[{\citenamefont{Asta et~al.}(2001)\citenamefont{Asta, Ozolins, and
  Woodward}}]{asta:fppheq}
\bibinfo{author}{\bibfnamefont{M.}~\bibnamefont{Asta}},
  \bibinfo{author}{\bibfnamefont{V.}~\bibnamefont{Ozolins}}, \bibnamefont{and}
  \bibinfo{author}{\bibfnamefont{C.}~\bibnamefont{Woodward}},
  \bibinfo{journal}{JOM - Journal of the Minerals Metals {\&} Materials
  Society} \textbf{\bibinfo{volume}{53}}, \bibinfo{pages}{16}
  (\bibinfo{year}{2001}).

\bibitem[{\citenamefont{Ceder et~al.}(2000)\citenamefont{Ceder, van~der Ven,
  Marianetti, and Morgan}}]{ceder:oxides}
\bibinfo{author}{\bibfnamefont{G.}~\bibnamefont{Ceder}},
  \bibinfo{author}{\bibfnamefont{A.}~\bibnamefont{van~der Ven}},
  \bibinfo{author}{\bibfnamefont{C.}~\bibnamefont{Marianetti}},
  \bibnamefont{and} \bibinfo{author}{\bibfnamefont{D.}~\bibnamefont{Morgan}},
  \bibinfo{journal}{Modelling Simul. Mater Sci Eng.}
  \textbf{\bibinfo{volume}{8}}, \bibinfo{pages}{311} (\bibinfo{year}{2000}).

\bibitem[{\citenamefont{Zunger}(1994)}]{zunger:NATO}
\bibinfo{author}{\bibfnamefont{A.}~\bibnamefont{Zunger}}, in
  \emph{\bibinfo{booktitle}{NATO ASI on Statics and Dynamics of Alloy Phase
  Transformation}}, edited by \bibinfo{editor}{\bibfnamefont{P.~E.}
  \bibnamefont{Turchi}} \bibnamefont{and}
  \bibinfo{editor}{\bibfnamefont{A.}~\bibnamefont{Gonis}}
  (\bibinfo{publisher}{Plenum Press}, \bibinfo{address}{New York},
  \bibinfo{year}{1994}), vol. \bibinfo{volume}{319}, p. \bibinfo{pages}{361}.

\bibitem[{\citenamefont{van~de Walle and Ceder}(2001)}]{vandeWalle02}
\bibinfo{author}{\bibfnamefont{A.}~\bibnamefont{van~de Walle}}
  \bibnamefont{and} \bibinfo{author}{\bibfnamefont{G.}~\bibnamefont{Ceder}},
  \bibinfo{journal}{J. Phase Equilib.} \textbf{\bibinfo{volume}{23}},
  \bibinfo{pages}{348} (\bibinfo{year}{2001}).

\bibitem[{\citenamefont{Ruban and Abrikosov}(2008)}]{Ruban08}
\bibinfo{author}{\bibfnamefont{A.~V.} \bibnamefont{Ruban}} \bibnamefont{and}
  \bibinfo{author}{\bibfnamefont{I.~A.} \bibnamefont{Abrikosov}},
  \bibinfo{journal}{Rep. Prog. Phys.} \textbf{\bibinfo{volume}{71}},
  \bibinfo{pages}{1} (\bibinfo{year}{2008}).

\bibitem[{\citenamefont{Jaynes}(2003)}]{Jaynes:prob}
\bibinfo{author}{\bibfnamefont{E.~T.} \bibnamefont{Jaynes}},
  \emph{\bibinfo{title}{Probability Theory: The Logic of Science (Vol. I)}}
  (\bibinfo{publisher}{Cambridge University Press},
  \bibinfo{address}{Cambridge, U.K.}, \bibinfo{year}{2003}).

\bibitem[{\citenamefont{Tipping}(2001)}]{Tipping01}
\bibinfo{author}{\bibfnamefont{M.~E.} \bibnamefont{Tipping}},
  \bibinfo{journal}{J. Machine Learn. Res.} \textbf{\bibinfo{volume}{1}},
  \bibinfo{pages}{211} (\bibinfo{year}{2001}).

\bibitem[{\citenamefont{Figueiredo and Jain}(2002)}]{Figueiredo02}
\bibinfo{author}{\bibfnamefont{M.~A.~T.} \bibnamefont{Figueiredo}} \bibnamefont{and}
  \bibinfo{author}{\bibfnamefont{A.~K.} \bibnamefont{Jain}},
  \bibinfo{journal}{IEEE Trans. Pattern Analys. Machine Intel.} \textbf{\bibinfo{volume}{24}},
  \bibinfo{pages}{381} (\bibinfo{year}{2002}).

\bibitem[{\citenamefont{Jones and Gunnarsson}(1989)}]{jones:ldareview}
\bibinfo{author}{\bibfnamefont{R.~O.} \bibnamefont{Jones}} \bibnamefont{and}
  \bibinfo{author}{\bibfnamefont{O.}~\bibnamefont{Gunnarsson}},
  \bibinfo{journal}{Rev. Mod. Phys.} \textbf{\bibinfo{volume}{61}},
  \bibinfo{pages}{689} (\bibinfo{year}{1989}).

\bibitem[{\citenamefont{Payne et~al.}(1992)\citenamefont{Payne, Teter, Allan,
  Arias, and Joannopoulos}}]{Payne:Pseudo-review}
\bibinfo{author}{\bibfnamefont{M.~C.} \bibnamefont{Payne}},
  \bibinfo{author}{\bibfnamefont{M.~P.} \bibnamefont{Teter}},
  \bibinfo{author}{\bibfnamefont{D.~C.} \bibnamefont{Allan}},
  \bibinfo{author}{\bibfnamefont{T.~A.} \bibnamefont{Arias}}, \bibnamefont{and}
  \bibinfo{author}{\bibfnamefont{J.~D.} \bibnamefont{Joannopoulos}},
  \bibinfo{journal}{Rev. Mod. Phys.} \textbf{\bibinfo{volume}{64}},
  \bibinfo{pages}{1045} (\bibinfo{year}{1992}).

\bibitem[{\citenamefont{Perdew et~al.}(1992)\citenamefont{Perdew, Chevary,
  Vosko, Jackson, Pederson, Singh, and Fiolhais}}]{perdew:appgga}
\bibinfo{author}{\bibfnamefont{J.~P.} \bibnamefont{Perdew}},
  \bibinfo{author}{\bibfnamefont{J.~A.} \bibnamefont{Chevary}},
  \bibinfo{author}{\bibfnamefont{S.~H.} \bibnamefont{Vosko}},
  \bibinfo{author}{\bibfnamefont{K.~A.} \bibnamefont{Jackson}},
  \bibinfo{author}{\bibfnamefont{M.~R.} \bibnamefont{Pederson}},
  \bibinfo{author}{\bibfnamefont{D.~J.} \bibnamefont{Singh}}, \bibnamefont{and}
  \bibinfo{author}{\bibfnamefont{C.}~\bibnamefont{Fiolhais}},
  \bibinfo{journal}{Phys Rev. B} \textbf{\bibinfo{volume}{46}},
  \bibinfo{pages}{6671} (\bibinfo{year}{1992}).

\bibitem[{\citenamefont{Sanchez et~al.}(1984)\citenamefont{Sanchez, Ducastelle,
  and Gratias}}]{Sanchez84}
\bibinfo{author}{\bibfnamefont{J.~M.} \bibnamefont{Sanchez}},
  \bibinfo{author}{\bibfnamefont{F.}~\bibnamefont{Ducastelle}},
  \bibnamefont{and} \bibinfo{author}{\bibfnamefont{D.}~\bibnamefont{Gratias}},
  \bibinfo{journal}{Physica} \textbf{\bibinfo{volume}{128A}},
  \bibinfo{pages}{334} (\bibinfo{year}{1984}).

\bibitem[{\citenamefont{Sluiter et~al.}(1996)\citenamefont{Sluiter, Watanabe,
  de~Fontaine, and Kawazoe}}]{Sluiter96}
\bibinfo{author}{\bibfnamefont{M.~H.~F.} \bibnamefont{Sluiter}},
  \bibinfo{author}{\bibfnamefont{Y.}~\bibnamefont{Watanabe}},
  \bibinfo{author}{\bibfnamefont{D.}~\bibnamefont{de~Fontaine}},
  \bibnamefont{and} \bibinfo{author}{\bibfnamefont{Y.}~\bibnamefont{Kawazoe}},
  \bibinfo{journal}{Phys. Rev. B} \textbf{\bibinfo{volume}{53}},
  \bibinfo{pages}{6137} (\bibinfo{year}{1996}).

\bibitem[{\citenamefont{D{\'{i}}az-Ortiz
  et~al.}(2007)\citenamefont{D{\'{i}}az-Ortiz, Dosch, and Drautz}}]{Drautz07}
\bibinfo{author}{\bibfnamefont{A.}~\bibnamefont{D{\'{i}}az-Ortiz}},
  \bibinfo{author}{\bibfnamefont{H.}~\bibnamefont{Dosch}}, \bibnamefont{and}
  \bibinfo{author}{\bibfnamefont{R.}~\bibnamefont{Drautz}},
  \bibinfo{journal}{J. Phys.: Condens. Matter} \textbf{\bibinfo{volume}{19}},
  \bibinfo{pages}{406206} (\bibinfo{year}{2007}).

\bibitem[{\citenamefont{Zunger et~al.}(2004)\citenamefont{Zunger, Wang, Hart,
  and Sanati}}]{Zunger04}
\bibinfo{author}{\bibfnamefont{A.}~\bibnamefont{Zunger}},
  \bibinfo{author}{\bibfnamefont{L.~G.} \bibnamefont{Wang}},
  \bibinfo{author}{\bibfnamefont{G.~L.~W.} \bibnamefont{Hart}},
  \bibnamefont{and} \bibinfo{author}{\bibfnamefont{M.}~\bibnamefont{Sanati}},
  \bibinfo{journal}{Model. Sim. Mater. Sci. Engr.}
  \textbf{\bibinfo{volume}{10}}, \bibinfo{pages}{685} (\bibinfo{year}{2004}).

\bibitem[{\citenamefont{Jansen and Popa}(2008)}]{Jansen08}
\bibinfo{author}{\bibfnamefont{A.~P.~J.} \bibnamefont{Jansen}}
  \bibnamefont{and} \bibinfo{author}{\bibfnamefont{C.}~\bibnamefont{Popa}},
  \bibinfo{journal}{Phys. Rev. B} \textbf{\bibinfo{volume}{78}},
  \bibinfo{pages}{085404} (\bibinfo{year}{2008}).

\bibitem[{\citenamefont{Mueller and Ceder}(2009)}]{Mueller09}
\bibinfo{author}{\bibfnamefont{T.}~\bibnamefont{Mueller}} \bibnamefont{and}
  \bibinfo{author}{\bibfnamefont{G.}~\bibnamefont{Ceder}},
  \emph{\bibinfo{title}{Bayesian approach to cluster expansions}}
  (\bibinfo{year}{2009}).

\bibitem[{\citenamefont{Levin et~al.}(2006)\citenamefont{Levin, Cockayne,
  Lufaso, Woicik, and Maslar}}]{Levin06}
\bibinfo{author}{\bibfnamefont{I.}~\bibnamefont{Levin}},
  \bibinfo{author}{\bibfnamefont{E.}~\bibnamefont{Cockayne}},
  \bibinfo{author}{\bibfnamefont{M.~W.} \bibnamefont{Lufaso}},
  \bibinfo{author}{\bibfnamefont{J.~C.} \bibnamefont{Woicik}},
  \bibnamefont{and} \bibinfo{author}{\bibfnamefont{J.~E.}
  \bibnamefont{Maslar}}, \bibinfo{journal}{Chem. Mater.}
  \textbf{\bibinfo{volume}{18}}, \bibinfo{pages}{854} (\bibinfo{year}{2006}).

\bibitem[{\citenamefont{Kresse and Joubert}(1999)}]{Kresse99}
\bibinfo{author}{\bibfnamefont{G.}~\bibnamefont{Kresse}} \bibnamefont{and}
  \bibinfo{author}{\bibfnamefont{J.}~\bibnamefont{Joubert}},
  \bibinfo{journal}{Phys. Rev. B} \textbf{\bibinfo{volume}{59}},
  \bibinfo{pages}{1758} (\bibinfo{year}{1999}).

\bibitem[{\citenamefont{Vanderbilt}(1990)}]{Vanderbilt:soft_pseudo}
\bibinfo{author}{\bibfnamefont{D.}~\bibnamefont{Vanderbilt}},
  \bibinfo{journal}{Phys. Rev. B} \textbf{\bibinfo{volume}{41}},
  \bibinfo{pages}{7892} (\bibinfo{year}{1990}).

\bibitem[{\citenamefont{Tikhonov}(1943)}]{tikhonov:reg}
\bibinfo{author}{\bibfnamefont{A.~N.} \bibnamefont{Tikhonov}},
  \bibinfo{journal}{Dokl. Akad. Nauk SSSR} \textbf{\bibinfo{volume}{39}},
  \bibinfo{pages}{195} (\bibinfo{year}{1943}).

\bibitem[{\citenamefont{Sundararaghavan and Zabaras}(2008)}]{Sundu08}
\bibinfo{author}{\bibfnamefont{V.}~\bibnamefont{Sundararaghavan}}
  \bibnamefont{and} \bibinfo{author}{\bibfnamefont{N.}~\bibnamefont{Zabaras}},
  \bibinfo{journal}{Phys. Rev. B} \textbf{\bibinfo{volume}{77}},
  \bibinfo{pages}{064101} (\bibinfo{year}{2008}).

\bibitem[{\citenamefont{Gale and Rohl}(2003)}]{Gale03}
\bibinfo{author}{\bibfnamefont{J.~D.} \bibnamefont{Gale}} \bibnamefont{and}
  \bibinfo{author}{\bibfnamefont{A.~L.} \bibnamefont{Rohl}},
  \bibinfo{journal}{Mol. Simul.} \textbf{\bibinfo{volume}{29}},
  \bibinfo{pages}{291} (\bibinfo{year}{2003}).

\end{thebibliography}
\end{document}